\documentclass[prb,twocolumn,showpacs,preprintnumbers,amsmath,amssymb,superscriptaddress]{revtex4}

\usepackage{graphicx}
\usepackage{dcolumn}
\usepackage{bm}
\usepackage{color}

\begin{document}

\title{Scanning tunneling spectroscopy study of $c$-axis proximity effect in epitaxial bilayer
manganite/cuprate thin films}

\author{I. Fridman}
\affiliation{Department of Physics, University of Toronto, 60 St. George St., Toronto, ON  M5S1A7, Canada}
 
\author{L. Gunawan}
\affiliation{Brockhouse Institute for Materials Research, McMaster University, 1280 Main Street West, Hamilton, ON, L8S4M1 Canada}

\author{G. A. Botton}
\affiliation{Brockhouse Institute for Materials Research, McMaster University, 1280 Main Street West, Hamilton, ON, L8S4M1 Canada}

\author{J. Y.T. Wei}
\affiliation{Department of Physics, University of Toronto, 60 St. George St., Toronto, ON  M5S1A7, Canada}
\affiliation{Canadian Institute for Advanced Research, Toronto, ON, M5G1Z8 Canada}


\begin{abstract}
Recent experimental studies have indicated novel superconducting proximity effects in thin-film heterostructures comprising ferromagnetic manganites and superconducting cuprates. To look for such effects microscopically, we performed scanning tunneling spectroscopy on La$_{2/3}$Ca$_{1/3}$MnO$_3$/YBa$_2$Cu$_3$O$_{7-\delta}$ (LCMO/YBCO) bilayer thin films.  $\emph{c}$-axis oriented films of varying thickness were grown on SrTiO$_{3}$ substrates using pulsed laser-ablated deposition. Heteroepitaxiality of the films was confirmed by cross-sectional transmission electron microscopy.  Tunneling spectra were measured at 4.2 K, and analyzed for signatures of a pairing gap on the LCMO layer.  For bilayer samples with LCMO thickness down to 5nm, asymmetric conductance spectra characteristic of single-layer LCMO films were observed, showing no clear gap structures.  These observations are consistent with a very short-range proximity effect involving spin-singlet pairs, and difficult to reconcile with longer-range proximity scenarios involving spin-triplet pairs.
\end{abstract}

\pacs{74.45.+c, 74.78.Fk, 74.50.+r}

\maketitle

\section{INTRODUCTION}
It is well known that ferromagnetism competes with spin-singlet superconductivity by favoring parallel alignment of electron spins.  In ferromagnet/superconductor (F/S) heterostructures, this competition gives rise to a variety of phenomena, such as a strong suppression of the superconducting proximity effect (PE) and the so-called $\pi$-state, where the superconducting order parameter penetrates into the ferromagnet with a spatially-modulated phase.  \cite{Buzdin2005,Kontos2001}  There has also been experimental evidence for spin-triplet superconductivity occurring in thin-film heterostructures made of a half-metallic ferromagnet and an $\emph{s}$-wave superconductor. \cite{Klapwijk2006}  Recent advances in thin-film growth of complex oxides have allowed such phenomena to be studied using the nearly half-metallic La$_{2/3}$Ca$_{1/3}$MnO$_3$ (LCMO) and the $\emph{d}$-wave superconductor YBa$_2$Cu$_3$O$_{7-\delta}$ (YBCO).  Epitaxial thin-film heterostructures of these lattice-compatible perovskites have shown several novel effects. Of particular interest is the dependence of the superconducting critical temperature $T_{c}$ on the LCMO layer thickness in LCMO/YBCO superlattices, showing rapid enhancement of $T_{c}$ below thicknesses that are substantially longer than the estimated PE depth.\cite{Sefrioui2003,Pang2004,Pena2004}   Similarly long length scales were also seen for the proximity-induced metal-to-insulator transition in LCMO/YBCO superlattices.\cite{Holden2004}  One proposed explanation for these effects is an anomalously long-ranged PE associated with spin-triplet pair formation in LCMO.\cite{Bergeret2005,Bergeret2001,Volkov2003,Halterman2007,Niu2007}  Unlike spin-singlet pairs, spin-triplet pairs are not easily broken by an exchange field, and can thus penetrate deep into the ferromagnet \cite{Buzdin2005}. Other experiments have indicated significant orbital reconstruction at the LCMO/YBCO interface\cite{Chakhalian2007} and nontrivial magnetic modulations in superlattices \cite{Stahn2005}, mechanisms that could also facilitate long-range superconducting PE in LCMO.

The superconducting PE in bilayer thin films can be directly probed with scanning tunneling spectroscopy (STS), by measuring the tunneling density of states (DOS) at known distances from the bilayer interface.  A proximity-induced gap in the DOS spectrum would be a signature of Cooper-pair formation and provide a measure of the pairing amplitude.\cite{vanson1987}  Previous studies have used STS to probe the PE of either a conventional or cuprate superconductor on a normal metal.\cite{Tessmer1996,Sharoni2004}  For F/S bilayers, STS has also been used to observe proximity-induced gap spectra on the surface of CuNi/Nb and SrRuO$_{3}$/YBCO thin films.\cite{Cretinon2005,Asulin2006,Asulin2007}  Since it is a local probe, STS is more immune to large-scale sample inhomogeneities that can affect bulk resistivity measurements, while being more sensitive to spatial variations in the quasiparticle DOS. 

In this paper, we present STS measurements at 4.2 K on epitaxially-grown LCMO/YBCO bilayer thin films as a function of the LCMO layer thickness.  Tunneling conductance spectra taken on LCMO/YBCO films with the LCMO layer thicker than 10 nm are similar to spectra taken on single-layer LCMO films, showing a $V$-shaped profile with no clear gap structure.  Bilayer films with the LCMO layer $\sim$ 5 nm thick show similar spectra with some variation versus tip position, but also no signature of a proximity-induced gap.  Our results indicate that the superconducting order parameter in the LCMO layer is suppressed well within $\sim$ 5 nm from the LCMO/YBCO interface, implying a very short-range PE consistent with predominantly spin-singlet pairing.

\begin{figure}
\includegraphics {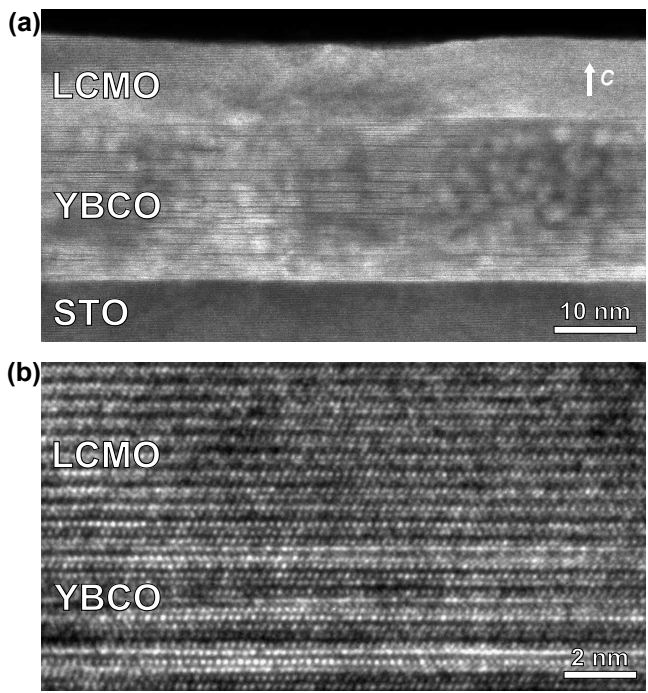}
\caption{\label{tem} (a) Transmission electron microscope cross-sectional image of a $\emph{c}$-axis oriented bilayer film with 10 nm of LCMO and 20 nm of YBCO grown on an STO substrate. The LCMO/YBCO interface is smooth over $\sim$ 100 nm and heteroepitaxial, as shown in the atomically resolved image (b).}
\end{figure}

\section{EXPERIMENT}

The LCMO/YBCO films used in our experiment were grown by pulsed laser-ablated deposition (PLD) on (001)-oriented SrTiO$_{3}$ (STO) substrates.  We used a 248 nm excimer laser, pulsed at 2-5 Hz, producing $\sim$2 J/cm$^{2}$ laser fluence on the target.  A layer of YBCO was first deposited on STO at 760$^{\circ}$C in 250 mTorr O$_{2}$, followed by \emph{in-situ} deposition of LCMO at 760$^{\circ}$C in 500 mTorr O$_{2}$.   We grew various films ranging in LCMO thickness from 5 to 20 nm and YBCO thickness from 20 to 40 nm.  All the films were confirmed to be superconducting by resistivity measurements, showing $T_{c}$ $\approx$ 84 and 89 K for films with LCMO thickness $\approx$ 5 and 10 nm, respectively.  

The microstructure of our LCMO/YBCO films was characterized by transmission electron microscopy (TEM).  Samples were prepared for cross-sectional imaging by mechanical polishing and dimpling prior to ion milling to electron transparency. TEM observations were carried out with a FEI Titan 80-300 (without an aberration corrector of the probe forming lens) in scanning mode using the high-angle annular angle dark field imaging method, which is sensitive to the atomic number contrast.  Figure \ref{tem} shows the cross-sectional TEM image of a LCMO/YBCO film, clearly indicating heteroepitaxial growth with unit-cell interfacial roughnesses over a range of $\sim$ 100 nm.  The roughness of the top surface is $\sim$ 3 nm.  Similar TEM images were seen on other LCMO/YBCO samples, attesting to the overall heteroepitaxiality of our PLD process, and providing strong evidence that our STS spectra were obtained on sufficiently smooth regions with good structural order.

Our STS measurements were made with a home-built scanning tunneling microscope (STM) operating in $\sim$ 1 mTorr of  $^{4}$He exchange gas at 4.2 K.  The LCMO/YBCO films were loaded into the STM immediately after growth, in order to minimize surface contamination.  Pt-Ir tips were used and the typical junction impedance was $\sim$10 G$\Omega$.  STS data were taken by suspending the STM feedback to fix the tip-sample distance and then measuring the tunneling current $I$ versus the bias voltage $V$ between sample and tip.  Fifty \emph{I-V} curves were averaged at each tip location and then numerically differentiated to yield the conductance \emph{dI/dV} spectra.  To ensure reproducibility of data, we obtained spectra at multiple locations over a scan range of 0.5 x 0.5 $\mu$m$^2$ on each sample, and measured several samples for each layer thickness.

\begin{figure}
\includegraphics {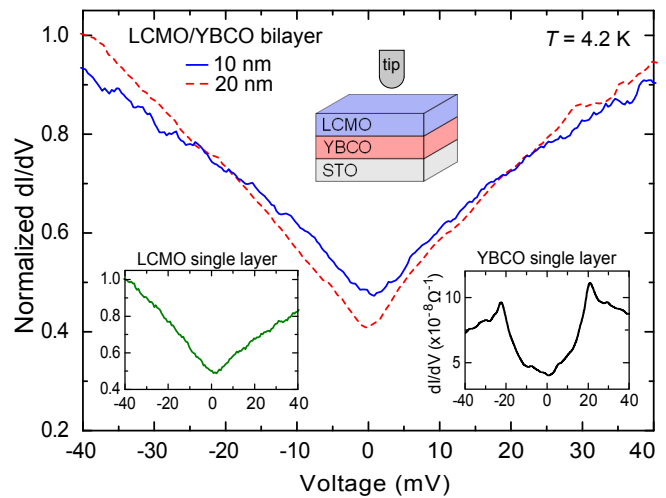}
\caption{\label{LCMO_comb}(Color online) Normalized conductance spectra measured on LCMO/YBCO bilayers with 10-nm-thick (blue solid curve) and 20-nm-thick (red dashed curve) LCMO, respectively. These spectra are similar to those measured on a single-layer LCMO film (left inset), but different from the features seen on a single-layer of YBCO grown under similar conditions (right inset, from Ref. \onlinecite{Ngai2007}). Top inset: experimental setup with the STM tip above a bilayer thin-film sample.}
\end{figure}

\section{RESULTS AND DISCUSSION}

First we present STS data measured on LCMO/YBCO films with the thickest LCMO layers.  The main plot of Figure \ref{LCMO_comb} shows the \emph{dI/dV} spectra for two films, one with 10 nm and the other with 20 nm of LCMO deposited over 20 nm of YBCO.  These plotted spectra have been normalized by their values
at 50 mV.  These spectra have a characteristic 'V' shape with some asymmetry, and substantial conductance at zero bias, indicating finite DOS at the Fermi level.  A clear gap structure is not seen, and the spectra are different from those measured on a single layer of superconducting YBCO grown under similar conditions by our group.\cite{Ngai2007}  These bilayer data are also qualitatively similar to data taken on single-layer LCMO films, as shown in the inset for a 60nm LCMO film deposited on STO substrate.  Similar V-shaped spectra have been seen in other transition-metal oxides, and appear to be characteristic of strongly-correlated oxide materials.\cite{Greene2001,Ray1995}  It is worth noting that in the case of the PE between Au and (001) YBCO films studied in Ref.~\onlinecite{Sharoni2004}, the proximity-induced superconducting gap decreases exponentially as a function of the distance from \emph{a}-axis facets at crystallite edges. In our experiments on films with 10-20 nm thick LCMO layers, we found no significant spectral variation with tip position, including near crystallite edges or over flat regions.  For all the spectra shown in Figure \ref{LCMO_comb}, no clear signatures of superconducting PE can be discerned within the characteristic V-shaped profile.  These observations are consistent with strong suppression of the PE in the LCMO layer. 

\begin{figure}
\includegraphics {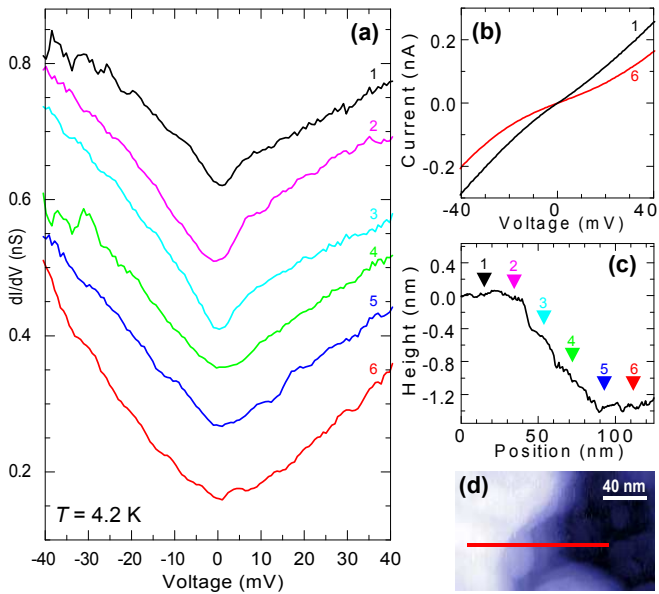}
\caption{\label{LCMO_5nm} (Color online) Spectral variation as a function of tip position measured on a LCMO/YBCO bilayer film with LCMO thickness of 5 nm.  (a) \emph{dI/dV} spectra measured in steps of 15 nm along a height gradient on the surface, starting at a high position (1) and ending at a low position (6).  Spectra are offset vertically for clarity. (b) \emph{I-V} curves for the characteristically metallic (1) and insulating (6) spectra. (c) Relative height of the surface measured using a topographic scan, with markers indicating the positions of the spectra measured in (a).  (d) Topographic image (1 V bias, 100 pA current) showing terraces with heights corresponding to an LCMO unit cell, with the measurement path indicated by a red line.}
\end{figure}

Next we present STS data for LCMO/YBCO films with thinner LCMO layers, and discuss the observed spectral dependence on tip position.  Figure \ref{LCMO_5nm} shows the \emph{dI/dV} spectra for a bilayer film with 5 nm of LCMO deposited over 40 nm of YBCO.  These spectra, shown vertically offset for clarity, were taken along a height gradient on the film, showing minor spectral variation.  As the tip is scanned from high to low regions, corresponding to bright and dark regions in the topography, the \emph{dI/dV} spectra broaden slightly in shape and the zero-bias conductance diminishes by a factor of 2.  We can identify two distinct types of spectra, i.e. a V-shaped type which is similar to spectra taken on bilayers with thicker LCMO, and another type which shows more rounding close to zero bias.  These two types of spectra correspond to linear and curved \emph{I-V} characteristics, which could be associated with metallic and insulating regions respectively, as were seen by previous STS measurements on single-layer LCMO films and attributed to electronic phase separation over nanometer length scales. \cite{Fath1999,Chen2003}  In these previous measurements, the LCMO films appear to become predominantly metallic well below the Curie temperature $T_M$ $\approx$ 260 K.  However, the electronic phase separation is also believed to be exacerbated by lattice strain,\cite{Mitra2005} thus making it more likely to appear in our bilayer films with very thin LCMO layers. It should be noted that the LCMO thickness in our films at each tip position could not be precisely determined, since their surface roughness was $\sim$ 3 nm.  Nevertheless, the STS characteristics of our LCMO/YBCO bilayer films with 5 nm of LCMO are generically similar to data taken on single-layer LCMO films, consistent with the LCMO layer not being superconducting.

Finally, we discuss the implications of our STS data on the superconducting PE in LCMO/YBCO heterostructures.  For conventional PE between a normal metal (N) and a spin-singlet superconductor (S), it is generally accepted that the pair potential penetrates into N within an exponential decay length $\xi_{N}$ which is typically $\sim$ 100 nm, while the pair potential is suppressed on the S side within a superconducting coherence length $\xi_{S}$.\cite{degennes}  For PE between a ferromagnetic metal (F) and S, it is believed that the pair-potential penetration is greatly diminished by the ferromagnetic exchange field, which also suppresses Andreev reflection at the F/S interface.\cite{deJong1995}  Thus the pair-potential decay length for a F/S junction should be far shorter than for an N/S junction.  For the F/S case, the proximity-induced pair potential in F is expected to oscillate and decay on a length scale of $\xi_{F0}=\hbar v_{F}/2E_{ex}$ and $\xi_{F}=\sqrt{\hbar D/2E_{ex}}$ in the clean and dirty limits, respectively, where $v_{F}$ is the Fermi velocity, $E_{ex}$ is the exchange energy, $D=v_F l/3$ is the diffusion coefficient and $l$ is the mean free path in F.\cite{Beasley1997, Buzdin2005}  For LCMO we estimate both $\xi_{F0}$ and $\xi_{F}$ to be $\approx$ 0.5 nm, by using $v_{F}$ = 7.4$\times10^{7}$ cm/s (Ref.~\onlinecite{Pickett1996}), $E_{ex}$ = 3 eV (Ref.~\onlinecite{Venky1998}) and $l$ $\sim$ a few unit cells (Ref.~\onlinecite{Cohn1997}).  These estimates indicate that proximity-induced superconductivity involving only spin-singlet pairs should be heavily suppressed in LCMO, decaying over just a few unit cells.  

Our STS results on c-axis LCMO/YBCO bilayer films are consistent with a short-ranged PE, which produces no clear gap structures in the DOS spectra down to LCMO thickness of $\sim$10 unit cells. This absence of \emph{microscopic} evidence for long-ranged PE in our bilayer films suggests that, in order to explain the very long ($\sim$10-100 nm) proximity length scales reported by \emph{macroscopic} measurements on manganite/cuprate superlattices,\cite{Sefrioui2003,Pang2004,Pena2004} more unconventional mechanisms would need to be considered. These could include orbital reconstruction
at the manganite/cuprate interface,\cite{Chakhalian2007} magnetic modulations in the superlattices,\cite{Stahn2005} and proximity coupling based on odd-frequency
pairing.\cite{Bergeret2005,Bergeret2001,Tanaka2007,Asano2007,Golubov2007} Here, we note that a recent STS study on $a$-axis-oriented LCMO/YBCO bilayers also reported a long proximity length scale.\cite{Millo2011} It may be helpful to examine the role of this $a$-axis proximity effect in $c$-axis superlattice samples in order to clarify the experimental difference between our local probe and the bulk measurements. Further studies are needed to elucidate any novel long-ranged physics that could be mediating the superconducting correlations in manganite/cuprate superlattices.

\section{CONCLUSION}

In summary, we have performed scanning tunneling spectroscopy at 4.2 K on LCMO/YBCO bilayer thin films.  Samples with LCMO layers 10 nm and thicker showed $V$-shaped spectra that are characteristic of single-layer LCMO films, without clear gap structures as would be expected for a proximity-induced pair potential.  Samples with LCMO layers down to 5 nm thickness showed some spectral variation with tip position but also no signatures of proximity-induced superconductivity.  Our results indicate that the proximity-induced pair potential in LCMO is suppressed well within 5 nm from the LCMO/YBCO interface, consistent with a very short-range F/S proximity effect involving spin-singlet pairs, and difficult to reconcile with longer-range proximity scenarios involving spin-triplet pairing.

\begin{acknowledgments}
This work was supported by grants from NSERC, CFI, OIT, OCE, and the Canadian Institute for Advanced Research under the Quantum Materials Program. GAB is grateful to NSERC for funds through the Discovery Grant program. TEM observations were carried out at the Canadian Centre for Electron Microscopy, a facility supported by NSERC and McMaster University.
\end{acknowledgments}


\end{document}